# Pluto's Atmosphere Does Not Collapse

C. B. Olkin[a]*, L. A. Young[a], D. Borncamp[a], A. Pickles[b], B. Sicardy[c], M. Assafin[d], F. B. Bianco[e], M. W. Buie[a], A. Dias de Oliveira[c,j], M. Gillon[f], R. G. French[g], A. Ramos Gomes Jr.[d], E. Jehin[f], N. Morales[h], C. Opitom[f], J. L. Ortiz[h], A. Maury[i], M. Norbury[b], F. B. Ribas[j], R. Smith[k], L. H. Wasserman[l], E. F. Young[a], M. Zacharias[m], N. Zacharias[m]

[a]Southwest Research Institute, Boulder, USA.

[b]Las Cumbres Observatory Global Telescope Network, Goleta, USA.

[c]Observatoire de Paris, Meudon, France.

[d]Universidade Federal do Rio de Janeiro, Observatorio do Valongo, Rio de Janeiro, Brazil

[e]Center for Cosmology and Particle Physics, New York University, New York, USA.

[f]Institut d'Astrophysique de I'Université de Liège, Liège, Belgium.

[g]Wellesley College, Wellesley, USA.

[h]Instituto de Astrofísica de Andalucía-CSIC, Granada, Spain

[i]San Pedro de Atacama Celestial Explorations (S.P.A.C.E.), San Pedro de Atacama, Chile

[j]Observatório Nacional / MCTI, Rio de Janeiro, Brazil

[k]Astrophysics Research Institute, Liverpool John Moores University, Liverpool, UK

[l]Lowell Observatory, Flagstaff, USA.

[m]United States Naval Observatory, Washington D. C., USA.

*Correspondence to: Catherine Olkin

**Keywords:** Pluto, atmosphere; Pluto, surface; Atmosphere, evolution; Occultations.

**Abstract**

Combining stellar occultation observations probing Pluto's atmosphere from 1988 to 2013, and models of energy balance between Pluto's surface and atmosphere, we conclude that Pluto's atmosphere does not collapse at any point in its 248-year orbit. The occultation results show an increasing atmospheric pressure with time in the current epoch, a trend present only in models with a high thermal inertia and a permanent $N_2$ ice cap at Pluto's north rotational pole.

**Introduction**

Pluto has an eccentric orbit, e=0.26, and high obliquity, 102-126° (Dobrovolskis & Harris, 1983), leading to complex changes in surface insolation over a Pluto year, and, therefore, in surface temperatures. When the first volatile ice species, $CH_4$, was discovered on Pluto's surface, researchers quickly recognized that these insolation and temperature variations would lead to large annual pressure variations, due to the very sensitive dependence of equilibrium vapor-pressure on the surface temperature. Pluto receives nearly three times less sunlight at aphelion than perihelion, prompting early modelers to predict that Pluto's atmosphere would expand and collapse over its orbit (Stern & Trafton, 1984). More sophisticated models were made in the 1990's (Hansen & Paige, 1996), after the detection of Pluto's atmosphere in 1988 and the discovery of $N_2$ as the dominant volatile in the atmosphere and on the surface. Similar models were run recently (Young, 2013), systematically exploring a range of parameter space. These models predict changes on decadal timescales, dependent on the thermal inertia of the substrate and the total $N_2$ inventory. Only in a subset of the models did pressures increase by a factor of two between 1988 and 2002, consistent with observations (Sicardy et al., 2003, Elliot et al. 2003). Continuing observations of Pluto's atmospheric pressure on decadal timescales constrain thermal inertia, providing insight into deeper layers of the surface that aren't visible in imaging.

**Observations**

Stellar occultations, where a body such as Pluto passes between an observer and a distant star, provide the most sensitive method for measuring Pluto's changing atmospheric pressure. Pluto was predicted to occult an R=14 star on May 4, 2013 (Assafin et al., 2010). This was one of the most favorable Pluto occultations of 2013 because of the bright star, slow shadow velocity (10.6 km/s at Cerro Tololo), and shadow path near large telescopes. An unusual opportunity to refine the predicted path of the shadow presented itself in March 2013 when Pluto passed within 0.5 arcsec of the occulted star six weeks before the occultation. The Portable High-Speed Occultation Telescope group (based at Southwest Research Institute, Lowell Observatory and Wellesley College) coordinated observations of the appulse from multiple sites including the 0.9 m astrograph at Cerro Tololo Inter-American Observatory (CTIO), the 1-m Liverpool Telescope on the Canary Islands, as well as the Las Cumbres Observatory Global Telescope Network (LCOGT) sites at McDonald Texas, CTIO Chile, SAAO South Africa, SSO Australia and Haleakala Hawaii. The appulse observations improved the knowledge of the shadow path location such that the final prediction was within 100 km of the reconstructed location. Occultation observations were obtained from the three 1.0-m LCOGT telescopes at Cerro Tololo (Brown, et al. submitted). The three telescopes have 1.0 m apertures and used identical instrumentation, an off-axis Finger Lakes Instrumentation MicroLine 4720 frame transfer CCD cameras, unfiltered. The cameras have a 2-second readout time, and autonomous observations were scheduled with different exposure times to provide adequate time resolution and minimize data gaps in the ensemble observation. We measured the combined flux from the merged image of Pluto, Charon and occultation star as a function of time using aperture photometry, and accounted for variable atmospheric transparency using differential photometry with five field stars. The light curves were normalized using post-occultation photometry of the field stars relative to the occultation star.

Observations were also attempted from the Research and Education Cooperative Occultation Network (RECON) from the western United States. This was an excellent opportunity to test the network and provided backup observing stations in case the actual path was further north than predicted. Observations were attempted at 14 sites and data were acquired at all sites, although in the end, all RECON sites were outside of the shadow path.

**Modeling**

In order to interpret an occultation light curve we need to have accurate knowledge of the precise location of the star relative to Pluto. The geometric solution is obtained by a simultaneous fit to 7 light curves from the following five sites: Cerro Burek Argentina, LCOGT at Cerro Tololo Chile, Pico dos Dias Brazil, La Silla Observatory Chile and San Pedro de Atacama Chile. The observation at San Pedro de Atacama was made using Caisey Harlingten's 0.5-m Searchlight Observatory Network Telescope. Details of the geometric solution will be given in a future paper. These sites span ~900 kilometers across the shadow covering more than 35% of Pluto's disk with chords both north and south of the centerline. The reconstructed impact parameter for LCOGT at Cerro Tololo is 370 ± 5 km with a mid time of 08:23:21.60 ± 0.05s UT on May 4.

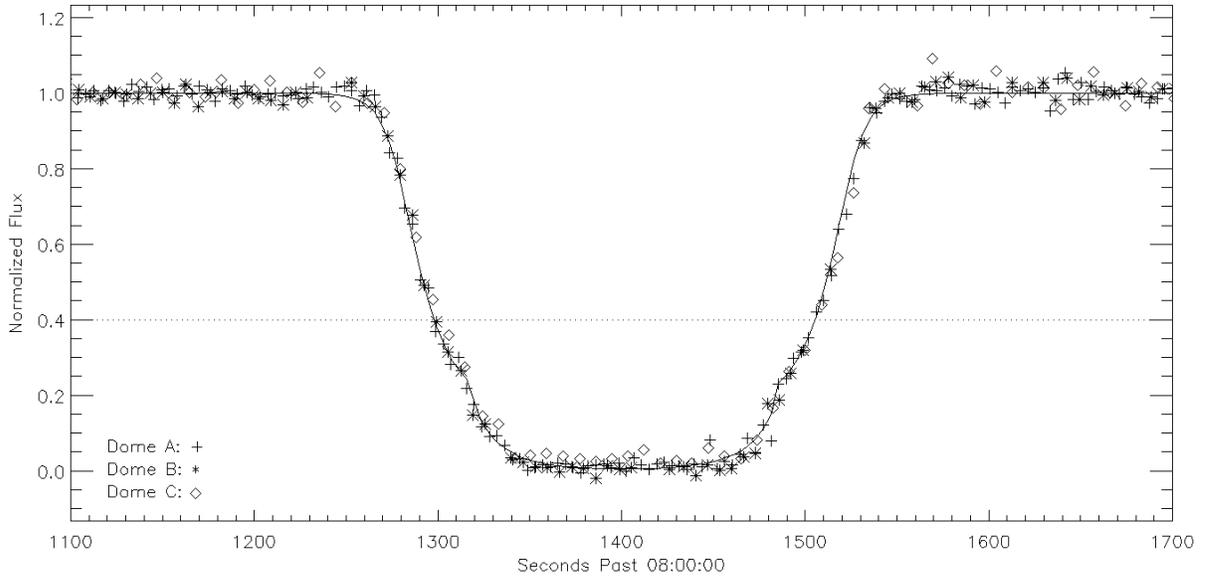

**Fig. 1**. The observed occultation light curves overlaid with the best fitting model. The line at normalized flux of 0.4 corresponds to 1275 km in Pluto's atmosphere. The transition from the upper atmosphere to the lower atmosphere occurs at a flux level of ~0.25 in these data. All three telescopes are 1.0-m telescopes located at the Cerro Tololo LCOGT node. The WGS 84 Coordinates of the three telescopes are (1) Dome A: Latitude: 30°.167383 S, Longitude: 70°.804789 W, (2) Dome B: Latitude: 30°.167331 S, Longitude: 70°.804661 W and (3) Dome C: Latitude: 30°.167447 S, Longitude: 70°.804681 W. All telescopes are at an altitude of 2201 m. The Dome A telescope used a 2-second integration time; Dome B used a 3-second integration time and Dome C used a 5-second integration time.

We fit the three LCOGT light curves simultaneously using a standard Pluto atmospheric model (Elliot & Young, 1992) that separates the atmosphere into two domains: a clear upper atmosphere with at most a small thermal gradient, and a lower atmosphere that potentially includes a haze layer. This model was developed after the 1988 Pluto occultation, which showed a distinct kink, or change in slope, in the light curve indicating a difference in the atmosphere above and below about 1215 km from Pluto's center. The lower atmosphere can be described with either a haze layer, or by a thermal gradient (Eshleman, 1989, Hubbard et al., 1990, Stansberry et al. 1994) or a combination of the two to match the low flux levels in the middle of the occultation light curves. We focus solely on the derived upper atmosphere parameters in this paper. Figure 1 shows the LCOGT light curves and the best fitting model with a pressure of 2.7 ± 0.2 microbar and a temperature of 113 ± 2 K for an isothermal atmosphere at 1275 km from Pluto's center. This atmospheric pressure extends the trend of increasing surface pressure with temperature since 1988.

Occultations are a sensitive measure of atmospheric pressure and have been a primary tool for detecting seasonal change on Pluto. Previous work (Young, 2013) combined stellar occultation observations from 1988 to 2010 and new volatile transport models to show that Pluto's seasonal variation falls into one of three classes: a class with high thermal inertia, which results in a northern hemisphere that is never devoid of $N_2$ ice (Permanent Northern Volatile, PNV, using the rotational north pole convention where the north pole is currently sunlit), a class with moderate thermal inertia and moderate $N_2$ inventory, resulting in two periods of exchange of $N_2$ ice between the northern and southern hemispheres that extend for decades after each equinox (Exchange with Pressure Plateau, EPP), and a class with moderate thermal inertia and smaller $N_2$ inventory, where the two periods of exchange of $N_2$ ice last only a short time after each equinox (Exchange with Early Collapse, EEC).

With this most recent stellar occultation of May 4 2013, we are able to distinguish between these three classes (Fig 2) of seasonal variation. The new data clearly preclude the EEC (Fig 2C) and EPP (Fig 2B) classes. Only the PNV class is consistent with the observations that show an increasing surface pressure in the current epoch. Both the EEC and EPP classes result in condensation of Pluto's atmosphere after solstice with surface pressures at the nanobar level or less (Young, 2013). The PNV model has a high thermal inertia, such that the atmosphere does not collapse over the course of a Pluto year with typical minimum values for the surface pressure of 10 microbar. At this surface pressure the atmosphere is collisional and present globally, and we conclude that Pluto's atmosphere does not collapse at any point during its 248-year orbit.

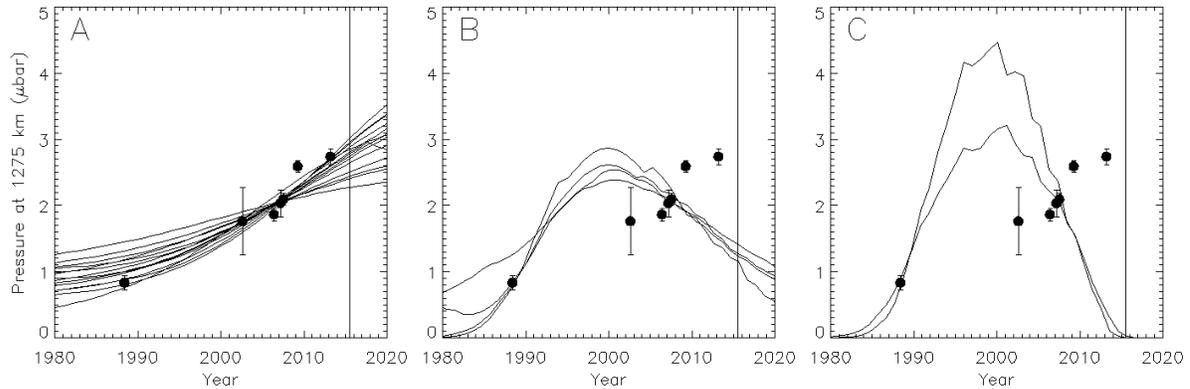

**Fig. 2.** Comparison of pressures derived from occultation measurements to pressures from volatile transport models (Young, 2013). Points indicate pressures in Pluto's atmosphere at 1275 km from Pluto's center, derived from fits of occultation data using the model of Elliot and Young (1992), and the points are repeated in each panel. These include six previously published measurements (Young, 2013) and the new measurement reported here. Lines correspond to modeled pressures for the Permanent Northern Volatiles (PNV) cases (2A), the Exchange with Pressure Plateau (EPP) cases (2B), and the Exchange with Early Collapse (EEC) cases (2C). The only class of model consistent with the increasing pressure from 1988 to 2013 is the Permanent Northern Volatile class. The vertical line in each panel is the closest approach of the New Horizons spacecraft to Pluto in July 2015.

**Discussion and Conclusions**

The PNV model runs that show an increasing atmospheric pressure with time over the span of stellar occultation observations (1988-2013) have thermal inertias of 1000 or 3162 J m$^{-2}$ s$^{-1/2}$ K$^{-1}$ (tiu). These values are much larger than the thermal inertia measured from daily variation in temperature on Pluto, ~18 tiu (Lellouch et al., 2011), or on other bodies such as Mimas, 16-66 tiu (Howett et al., 2011). The range of thermal inertias derived for Pluto from this work is comparable to that for pure, non-porous $H_2O$ ice at 30-40 K, 2300 to 3500 tiu (Spencer & Moore, 1992). This points to a variation of thermal inertia with depth. The variation of temperature over a day probes depths of ~1 m, while the seasonal models depend on conditions near 100 m, indicating that the thermal inertia is lower near the surface (~1 m) than at depth (~100 m). Evidence for large thermal inertias at the depths probed by seasonal variation has also been seen on Triton. Models that best explain the presence of a $N_2$ cap on the summer hemisphere of Triton during the 1989 Voyager encounter have thermal inertias greater than 1000 tiu (21). Also large-thermal inertia models for Triton (Spencer & Moore, 1992) are further supported by the large increase in atmospheric pressure observed on Triton from 1989 to 1995 (Olkin et al., 1997, Elliot et al., 2000). Pluto and Triton are similar in size, density and surface composition. They may also be similar in their substrate thermal inertia properties.

Pluto's atmosphere is protected from collapse because of the high thermal inertia of the substrate. The mechanism that prevents the collapse is specific to Pluto, because it relies on Pluto's high obliquity and the coincidence of equinox with perihelion and aphelion. In the PNV model, volatiles are present on both the southern and northern hemispheres of Pluto just past aphelion. Sunlight absorbed in the southern hemisphere (the summer hemisphere from aphelion to perihelion) powers an exchange of volatiles from the southern hemisphere to the northern (winter) hemisphere. Latent heat of sublimation cools the southern hemisphere and warms the northern hemisphere, keeping the $N_2$ ice on both hemispheres the same temperature. This exchange of volatiles continues until all the $N_2$ ice on the southern hemisphere sublimates and is condensed onto Pluto's northern hemisphere. Once this occurs at approximately 1910 in Fig. 3, the northern (winter at this time) hemisphere is no longer warmed by latent heat, and begins to cool. However, the thermal inertia of the substrate is high, so the surface temperature on the northern hemisphere does not cool quickly. The ice temperature drops by only a few degrees K before the $N_2$-covered areas at mid-northern latitudes receive insolation again, in the decades before perihelion, as shown in Fig. 3 from 1910 to 1970.

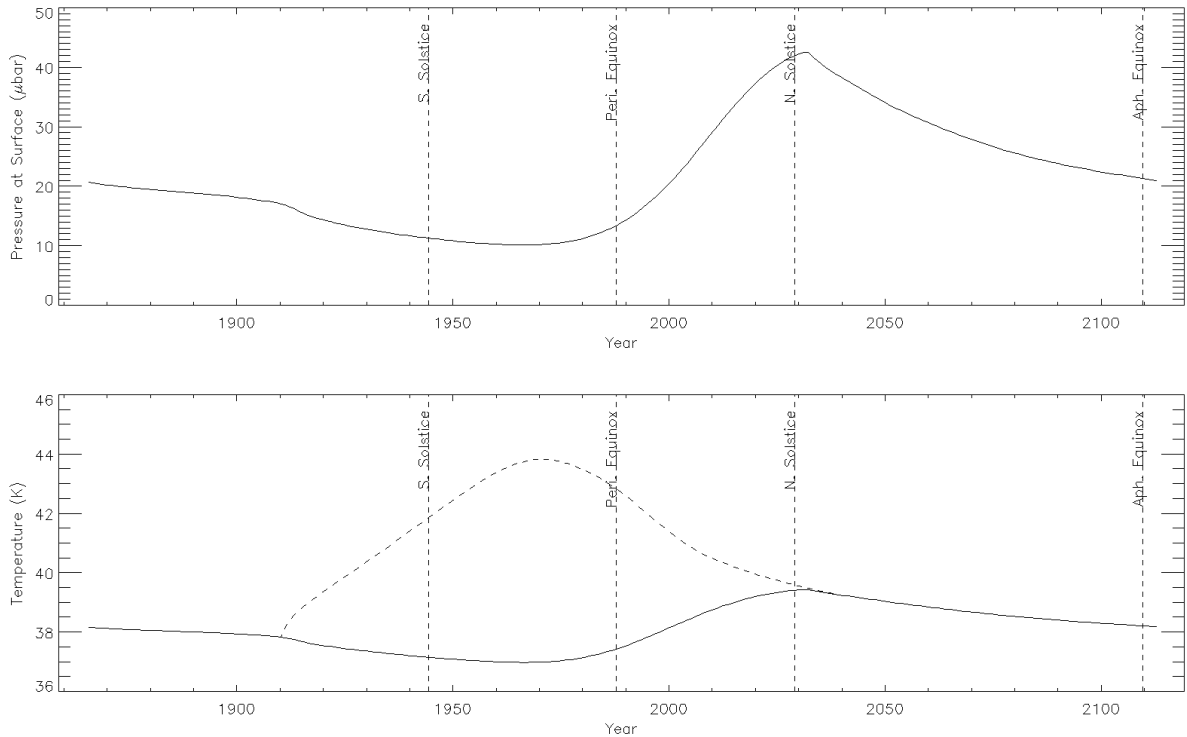

**Fig. 3.** The surface pressure (top panel) and temperature (lower panel) over a Pluto year for a typical PNV case. The surface pressure reaches a minimum of ~10 microbar. The temperature of the $N_2$ ice (solid line) and of a mid-southern latitude (-60°, dashed line) are indicated. At any given time, all the $N_2$ ice on Pluto's surface is at the same temperature due to the transfer of energy from condensation and sublimation. Bare, $N_2$-ice free regions can have temperatures higher than the ice temperature, as seen from 1910 to 2030 above. Southern solstice, equinox at perihelion, northern solstice, and equinox at aphelion are indicated for the current Pluto year.

Near the perihelion equinox, the southern hemisphere surface is warm, ~42 K, because the $N_2$-free substrate was illuminated for the preceding eight decades (1910-1990, in Fig. 3). Approaching and after equinox (at perihelion), the southern hemisphere receives less sunlight, and radiatively cools slowly due to high thermal inertia. Once the surface cools to the $N_2$ ice temperature (in approximately 2035, see Fig 3), the $N_2$ gas in the atmosphere will condense onto the southern hemisphere, and there begins a period of exchange transferring $N_2$ from the summer (northern) hemisphere to the winter (southern) hemisphere. However, this period of flow lasts only until equinox at aphelion. The period of exchange is not long enough to denude the northern hemisphere, thus as Pluto travels from perihelion to aphelion, $N_2$ ice is always absorbing sunlight on the northern hemisphere keeping the ice temperatures relatively high throughout this phase and preventing collapse of Pluto's atmosphere.

The PNV model is testable in multiple ways. In 2015, The New Horizons spacecraft will fly past Pluto providing the first close-up investigation of Pluto and its moons (Stern et al., 2008; Young et al. 2008). The infrared spectrometer on New Horizons will map the composition across Pluto's surface. We will be able to compare the $N_2$ ice distribution predicted by the Permanent Northern Volatile model with the observed ice distribution determined by New Horizons. The REX instrument on New Horizons will provide thermal measurements to compare with the surface temperatures predicted by the PNV models. From the UV solar and stellar occultations of Pluto, the New Horizons mission will determine the composition of Pluto's atmosphere as well as the thermal structure in the thermosphere. From the Radio Science experiment, the pressure and temperature profiles in Pluto's lower atmosphere will be determined. All of these data provide a test of the PNV model.

In addition to this close-up comprehensive investigation of Pluto by the New Horizons spacecraft, our conclusion that Pluto's atmosphere does not collapse can be tested by regular stellar occultation observations from Earth. The current epoch is a time of significant change on Pluto. Most of the PNV models show a maximum surface pressure between 2020 and 2040.

Regular observations over this time period will constrain the properties of Pluto's substrate and the evolution of its atmosphere.

**Acknowledgments**

This work was supported in part by NASA Planetary Astronomy grant NNX12AG25G. The Liverpool Telescope is operated on the island of La Palma by Liverpool John Moores University in the Spanish Observatorio del Roque de los Muchachos of the Instituto de Astrofisica de Canarias with financial support from the UK Science and Technology Facilities Council.

**References**

Assafin, M., *et al.*, Precise predictions of stellar occultations by Pluto, Charon, Nix, and Hydra for 2008-2015. *Astron. Astrophys.* **515** A32, 1-14 (2010).

Brown, T. M., *et al*., Las Cumbres Observatory Global Telescope Network. *PASP*, submitted.

Dobrovolskis, A. R., A. W. Harris, The obliquity of Pluto. *Icarus* **55** 231-235 (1983).

Elliot, J. L., L. A. Young, Analysis of stellar occultation data for planetary atmospheres. I. model fitting, with application to Pluto. *Astron. J.* **103**, 991-1015 (1992).

Elliot, J. L., *et al*., The prediction and observation of the 1997 July 18 stellar occultation by Triton: more evidence for distortion and increasing pressure in Triton's atmosphere. *Icarus* **148** 347-369 (2000).

Elliot, J. L., *et al.*, The recent expansion of Pluto's atmosphere. *Nature* **424** 165-168. (2003).

Eshleman V. R., Pluto's atmosphere: Models based on refraction, inversion, and vapor-pressure equilibrium. *Icarus* **80** 439-443 (1989).

Hansen, C. J., D. A. Paige, Seasonal nitrogen cycles on Pluto. *Icarus* **120** 247-265 (1996).

Howett, C. J. A., *et al.*, A high-amplitude thermal inertia anomaly of probably magnetospheric origin on Saturn's moon Mimas. *Icarus* **216** 221-226 (2011).

Hubbard, W. B., R. V. Yelle, J. I. Lunine, Nonisothermal Pluto atmosphere models. *Icarus* **84** 1-11 (1990).

Lellouch, E., J. Stansberry, J. Emery, W. Grundy, D. P. Cruikshank, Thermal properties of Pluto's and Charon's surfaces from Spitzer observations. *Icarus* **214** 701-716 (2011).

Olkin, C. B., *et al.*, The thermal structure of Triton's atmosphere: results from the 1993 and 1995 occultations. *Icarus* **129** 178-201 (1997).

Sicardy, B., *et al.*, Large changes in Pluto's atmosphere as revealed by recent stellar occultations. *Nature* **424** 168-170 (2003).

Spencer, J. R., J. M. Moore, The influence of thermal inertia on temperatures and frost stability on Triton. *Icarus* **99** 261-272 (1992).

Stansberry, J. A., J. I. Lunine, W. B. Hubbard, R. V. Yelle, D. M. Hunten, Mirages and the nature of Pluto's atmosphere. *Icarus* **111** 503-513 (1994).

Stern, S. A., L. Trafton, Constraints on bulk composition, seasonal variation, and global dynamics of Pluto's atmosphere. *Icarus* **57** 231-240 (1984).

Stern, S. A., *et al.*, The New Horizons Pluto Kuiper belt mission: an overview with historical context. *Sp. Sci. Rev.* **140** (2008).

Young, L. A., *et al.*, New Horizons: anticipated scientific investigations at the Pluto system. *Sp. Sci. Rev.* **140** 93-127 (2008).

Young, L. A., Pluto's Seasons: new predictions for New Horizons. *Ap. J. L.* **766** L22-L28 (2013).